\documentclass[proof]{WileyASNA-v1}

\articletype{Article Type}%

\received{1 October 2020}
\revised{}
\accepted{}

\begin{document}

\title{Stellar Initial Mass Function over a range of redshifts}

\author[1]{Rafeel Riaz*}

\author[1]{Dominik R.G. Schleicher}

\author[2]{Siegfried Vanaverbeke}

\author[3,4]{Ralf S. Klessen}

\authormark{Rafeel Riaz}

\address[1]{Departamento de Astronom\'ia, Facultad Ciencias F\'isicas y Matem\'aticas, Universidad de Concepci\'on, Av. Esteban Iturra s/n Barrio, Universitario, Casilla $160$-C, Concepci\'on, Chile}

\address[2]{Centre for mathematical Plasma-Astrophysics, Department of Mathematics, KU Leuven, Celestijnenlaan 200B, 3001 Heverlee, Belgium}  

\address[3]{Universit{\"a}t Heidelberg, Zentrum f{\"u}r Astronomie, Institut  f{\"u}r Theoretische Astrophysik, Albert-Ueberle-Str. 2,  69120 Heidelberg, Germany} 

\address[4]{Universit{\"a}t Heidelberg, Interdisziplin{\"a}res Zentrum f{\"u}r Wissenschaftliches Rechnen, Im Neuenheimer Feld 205,  69120 Heidelberg, Germany}  

\corres{*Departamento de Astronom\'ia, Universidad de Concepci\'on, Chile.  \email{rriaz@astro-udec.cl}}

\abstract{The stellar Initial Mass Function (IMF) seems to be close to universal in the local star-forming regions. However, this quantity of a newborn stellar population responds differently at gas metallicities $Z \sim$ $Z_\odot$~ than $Z$ = 0. A view on the cosmic star formation history suggests that the cooling agents in the gas vary both in their types and molecular abundances. For instance, in the primordial gas environment, the gas temperature can be higher by a factor of 30 as compared to the present day. Stellar radiation feedback and cosmic microwave background (CMB) radiation may even contribute towards increasing the floor temperature of the star-forming gas which subsequently can leave profound impacts on the IMF. We present the contribution of the radiation sources towards the thermodynamical evolution of the Jeans unstable gas cloud which experiences fragmentation and mass accretion. We find evidence which suggests that the latter becomes the dominant process for star formation efficiencies (SFE) above 5 - 7 \%, thus increasing the average mass of the stars. We focus on the isolated and binary stellar configurations emerging during the gas collapse. The binary fraction on average remains 0.476 and contributes significantly towards the total SFE of 15 \%. }

\keywords{hydrodynamics, numerical method, protostars, accretion, IMF}

\maketitle

\section{The Introduction}

First realized in the mid of $20^{th}$ century by \citet{Salpeter1955} the stellar initial mass function (IMF) remained one of the most deeply investigated features in the realm of star formation \citep{Applebaum2020, Chabrier2005}. In general, it describes the ratio of the low-mass stars to the high mass stars in any newborn stellar population. It has enabled us to grasp the attribute of nature in which stellar masses are assigned a typical trend for a given stellar mass distribution. The IMF has been found to be almost universal within the Milky Way galaxy and in the nearby star forming regions \citep{chabrier2003galactic, Kroupa2001}. However, studies have also revealed that there exists a fair chance that it may not be entirely universal throughout the Cosmos \citep{Kroupa2019, Bastian2010}. Specially, when the cosmic star-formation history is taken into account. There can be multiple factors such as primordial cooling agents H$_{2}$ and HD, various levels of gas metallicity, presence of cosmic dust, cosmic rays, and the cosmic microwave background (CMB) radiation which may affect the star forming environments. These factors with their varying strengths of influence over a wide range of redshifts ($z$) may introduce distinctive trends in the IMF and can pose a challenge to its universality. Starting from the star-formation epoch some 200 Myr after the Big Bang, numerical studies have shown that due to the less efficient cooling agents present in the primordial gas massive stars are the most likely outcome of any local gas collapse in the early universe \citep{klessen2019formation, Aoki2014,Limongi2012,Silk2006}. This is unlike what we observe in the present day star-forming gas which is more enriched with metals and dust that collectively provide efficient cooling mechanisms \citep{Glover2012,Schneider2006,Hartmann2001}. This results in significantly low Jeans masses that during their thermodynamical evolution give birth to low-mass stars. Cosmic rays with the ability to permeate the gas throughout the entire galaxy even up to the coolest part of the interstellar medium \citep{Steinhardt2020} can be influential in regulating the thermal state of the gas. Similarly, the CMB radiation can be quite influential in regulating the thermal states of star forming gas and hence setting up the temperature floor \citep{Schneider2010,Bailin2010}. In this paper we discuss the role of such possible radiation sources in the gas clouds and analyze their subsequent effects on the resulting stellar masses appear in the star forming gas. We also focus on how the isolated and binary stellar configurations appear with their mass characteristics. For the structure of this paper, we describe the methodology of our simulations in section 2 which is followed by the discussion on the results in section 3. In section 4 we present our conclusions.

\section{Methodology}\label{sec2}

Our numerical scheme is known as smoothed particle hydrodynamics (SPH). We use the computer code GRADSPH \footnote{Webpage GRADSPH: http://www.swmath.org/software/1046} \citep{vanaverbeke2009gradsph}.
We set up our models as the total mass of 30 $M_{\odot}$ inside the spherical gas cloud represented by a total of 250025 SPH particles with the cloud radius of 0.168 pc. The initial gas density is $\rho_{i} = 1.0 \times 10^{-19}$~g cm$^{-3}$. At that density, assuming a metallicity of at least 0.1 solar, the gas may evolve in an approximately isothermal manner, as demonstrated e.g. by \citet{Omukai2005} and \citet{Grassi2017}. In this manuscript, we are investigating only the approximate trends, and will therefore assume an initially isothermal evolution, with a transition to the adiabatic regime once the gas becomes optically thick. 
Our equation of state is barotropic with the following form,
\begin{equation} \label{EOS}
P=\rho c_{0}^{2}\left[1+\left(\frac{\rho}{\rho_{\rm crit}}\right)^{\gamma-1}\right].
\end{equation}
To derive the values for the critical density \textit{$\rho$}$_{\rm crit}$ at which the phase transition from isothermal to adiabatic takes place in each of our models we use equation 20 of \citet{Omukai2005}. We consider the balance between the cooling which is dominated by continuum emission via dust thermal emission and compressional heating, implying  

\begin{equation} \label{transition}
T=\left(\frac{k^{3}}{12\sigma^{2}m_{\rm H}}\right) ^{1/5} {n^{2/5}_{\rm H}},
\end{equation}
where $k$, $\sigma$, $m_{\rm H}$, and $n_{\rm H}$ are the Boltzmann constant, Stefan-Boltzmann constant, mass of the hydrogen atom, and the number density of the gas, respectively. For the mass density, we have
\begin{equation} \label{transition}
\rho=n\mu m_{\rm H},
\end{equation}
which leads to the following expression to determine the critical density \textit{$\rho$}$_{crit}$:
\begin{equation} \label{transition}
\rho_{\rm crit}=6.115\times10^{-16}~T^{5/2}.   
\end{equation}

The relation $T_{\rm r}$ = 2.725~(1 + $z$) is used for our selected range of temperatures which approximately corresponds to the redshift $z$ = 2.7$-$17.3, if interpreted to correspond to the temperature of the CMB (see Table 1).
For all models considered in the present work the gas is initially isothermal. The free fall time for considered models is 1.888 Myr for the standard initial condition defined here.
The code uses internal dimensionless units which are defined by setting $G=M=R=1$.

We inject a spectrum of incompressible turbulence into the initial conditions in our models. Two distinct values for the initial seed while creating a velocity distribution for the injected turbulence are followed in our models as $set~1$ (M1a - M5a) and $set~2$ (M1b - M5b). The gas in each model is initially transonic. The initial condition is also characterized by the parameters $\alpha$ and $\beta$, which correspond to the ratio of thermal and kinetic energies to the gravitational potential energy of the system, respectively. The value of $\alpha$ changes according to the initial temperature set in each model (see Table 1) while the value of $\beta$ remains 0.02 in all models. The gas sphere is in solid-body rotation with angular rotational frequency $\omega$ = 2.912 x 10$^{-14}$ rad s$^{-1}$. Our treatment of sink particles (protostars) takes advantage of the introduced sink particle algorithm for SPH calculations described by \citet{Price2017}. We also consider the merger of sink particles which is also taken into account by \citet{stacy2013}. Our minimum resolvable mass inside the simulation domain is $M_{\rm resolvable} = 1.199 \times 10^{-2}$~$M_\odot$. A constant accretion radius $r_{\rm acc}$ = 1 au for the sink particles is used, and $r_{\rm acc}$ always remains greater than the Jeans length in our simulations.

The binarity, specially for the close binary systems with extreme mass ratios can be one of the major observational constraints in estimating the true stellar luminosities and hence determining the correct masses of the stars. However, this can be studies in numerical simulations, and in the current study we consider the mass spectrum of the newborn stellar population and analyze the single and binary star systems individually. We aim to explore the effect of a mean change of gas temperature with Cosmic redshift, assuming otherwise similar cloud properties. We focus particularly on trends regarding the IMF. Our initial setup of each molecular cloud is in the form of higher density gas which is coupled to the dust and making the dust cooling relevant. We terminate all of our simulations when the ratio of envelope mass to initial cloud mass ($M_{\rm env}/M_{\rm c}$) becomes 85\%, and we utilize the visualizing tool SPLASH which is publicly available to the community by \citet{price2007splash}. 

\section{Results and Discussion}

We present and then summarize in Table 1 our five models with two different seeds and their entire evolution. There the overall outcome is presented when the SFE ($\xi$) in collapsing gas clouds reaches $\xi$ = 15 \% mark. Figure 1 shows the collapsed state of each model when rotating gas cloud forms disk structure which subsequently experiences the process of fragmentation giving birth to the protostars. The first and second columns in this figure from top to bottom depict respectively the models which are constituted by the two seed sets ($set~1$ and $set~2$) described before. Moving from initially the coldest gas cloud to the warmest, both the disk morphology and the strength of fragmentation in it seem to follow a trend, where the former exhibits more extended disk with greater number of fragments while the latter shows relatively confined disk with lesser number of fragments. This feature of disk fragmentation has the potential to affect profoundly the stellar IMF. Based upon this we suspect that the conditions in higher redshift environments where the external radiation sources such as the CMB radiation set a higher floor temperature in the star forming gas, favor a top-heavy IMF. As the universe gets older the CMB radiation drops down due to space-time expansion and only manages to contribute towards setting up the gas temperature which primarily in the presence of dust more controlled by the molecules such as CO and H$_{2}$ \citep{glover2012star,tress2020simulations}. This ever decreasing role of at least the CMB radiation may result in a more bottom-heavy IMF since the Jeans mass remains relatively low in such environments.        

\begin{figure}[t]
	\centerline{\includegraphics[width=80mm,height=27.4pc]{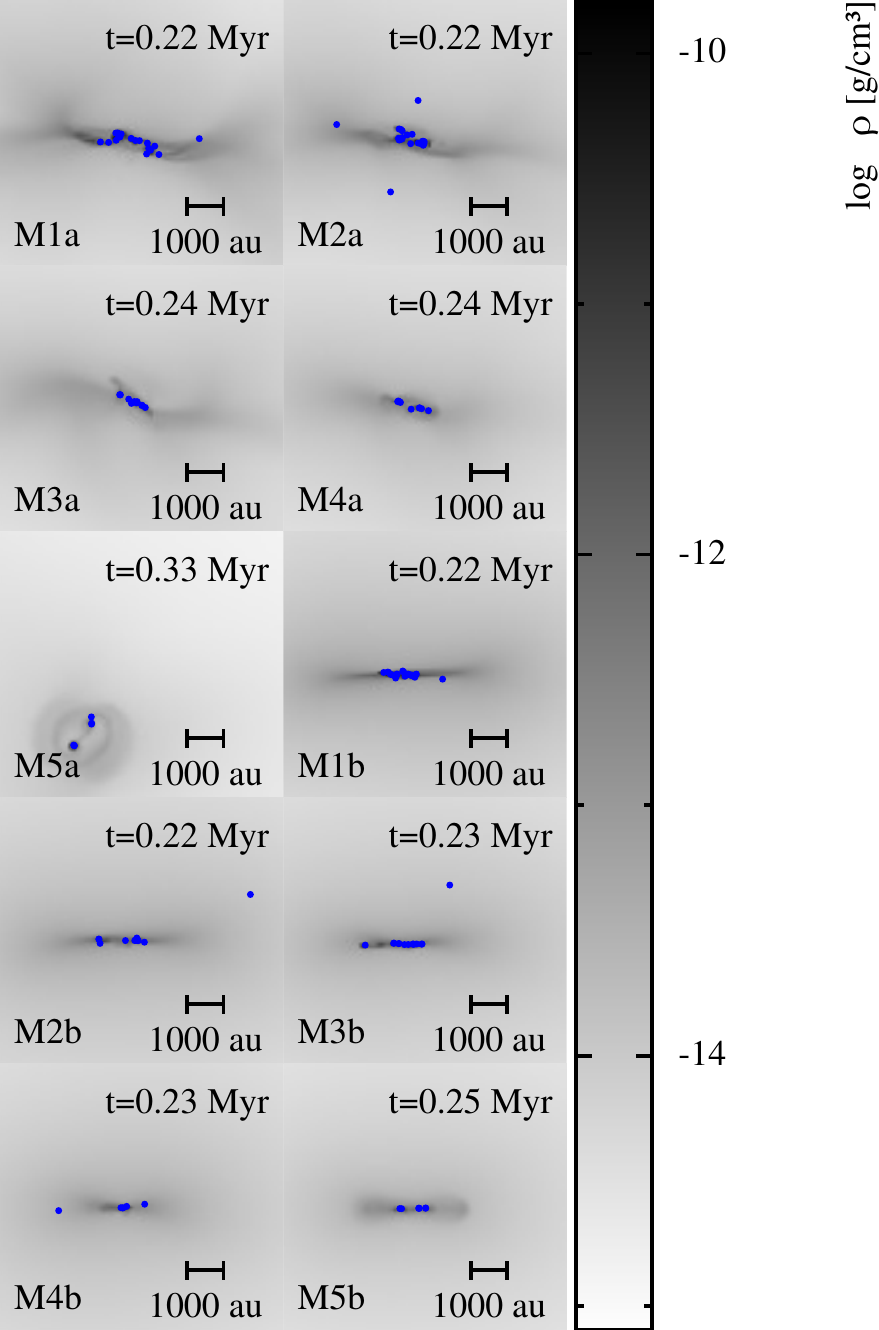}}
	\caption{Simulation results for models M1a$-$M5a and models M1b$-$M5b at the end of the computed evolution of each model following the seed 1 and seed 2, respectively. Each panel shows an edge-on view of the density in the xz-plane. The shaded bar on the right shows log ($\rho$) in g cm$^{-3}$. The corresponding dynamical time in Myr is shown at the top-right of each panel. Each calculation was performed with 250025 SPH particles. Also, these snapshots represent stages of evolution where star formation efficiency (SFE) in each model reaches $\xi$ = 15 \%. \label{Fig1}}
\end{figure}

Figure 2 shows the history of formation of the protostars inside the collapsing gas clouds described in our two sets of models M1a$-$M5a and M1b$-$M5b. The top and the bottom rows cover the first and second seed case, respectively. Every spike is a manifestation of additional number of protostars which form as the gas cloud evolves thermodynamically \citep{boss1980protostellar}. This number also declines when protostellar mergers happen inside the collapsing gas.  It remains evident that cold gas clouds during their evolution keep protostar formation throughout their history. Whereas, in warm gas clouds there are fewer protostars forming. In the later part of the evolution the absence of spikes is an indication that the gas keeps evolving without giving many births to the new protostars inside the adiabatic phase of collapse. This primarily is due to the higher Jeans mass condition in warmer gas clouds which restricts new clumps to form and become self-gravitating objects \citep{bate2005dependence}. Subsequently, the warmer gas clouds take relatively more time than their counterparts to reach to the stage of $\xi$ = 15 \% (a criterion that we follow to terminate our model simulations). In general, the formation history of protostars remains distinct for cold and warm gas environments. This may leave a profound impact on the IMF as the latter can contribute to the building up of a top-heavy IMF in star forming regions at higher redshifts.

\begin{center}
	\begin{table*}[t]%
		\caption{Summary of the two sets of models M1a$-$M5a with the first random seed and M1b$-$M5b with the second random seed (corresponding to two different realizations of turbulence, with the same statistical properties). The table is constructed at the time  when SFE ($\xi$) reaches 15 \% in each model. The table describes the ratio of kinetic to gravitational potential energy of the cloud ($\alpha$), the initial temperature ($T_{\rm i}$), the corresponding cosmic redshift, the Jeans mass of the cloud evaluated at the initial gas density ($M_{\rm J}$), the total number of protostars produced ($N_{\rm max}$), the final number of protostars after mergers ($N_{\rm proto}$), the final number of binaries ($N_{\rm binary}$), binary contribution towards SFE ($\xi_{\rm binary}$),  the mean mass of the total protostellar population ($M_{\rm mean}$) with its variance. The error estimation is performed with a confidence interval of 68.3 \%.\label{tab1}}
		\centering
		\begin{tabular}{cccccccccccc}
			\hline
			Model & $\alpha$ & $T_{\rm i}$ (K)& Redshift ($z$) & $M_{\rm J}$ ($M_{\odot}$) & $N_{\rm max}$ & $N_{\rm proto}$ & $N_{\rm binary}$ & $\xi_{\rm binary}$ (\%)   & $M_{\rm mean}$ ($M_{\odot}$)  \\
			\hline
			M1a  & 0.115     & 10  &2.66 &   4.855 & 26 & 23 & 3 &11.80&  0.157$\pm$ 0.0574   \\
			M2a  & 0.231     & 20  &6.33&   9.710 & 31 & 30 & 6 &14.30&  0.1512$\pm$ 0.0437   \\
			M3a  & 0.347	 & 30  &10.0&  14.565 & 17 & 12 & 3 &14.88& 0.3750$\pm$ 0.1130    \\
			M4a  & 0.463     & 40  &13.67&  19.420 & 23 & 9 & 2 &10.97& 0.5000$\pm$ 0.1050   \\
			M5a  & 0.578     & 50  &17.34&  24.275 & 12 &  9 & 2 &13.03& 0.5000$\pm$ 0.1430   \\
			M1b  & 0.115     & 10  &2.66 &   4.855 & 28 & 24  & 3 &12.73& 0.1875$\pm$ 0.0420    \\
			M2b  & 0.231     & 20  &6.33&   9.710 & 20 & 12 & 3 &14.85&0.3750$\pm$ 0.1070   \\
			M3b  & 0.347	 & 30  &10.0&  14.565 & 19 & 14 & 4 &13.21&0.3214$\pm$ 0.1020  \\
			M4b  & 0.463     & 40  &13.67&  19.420 & 31 & 15 & 1 &14.85&0.3000$\pm$ 0.1060  \\
			M5b  & 0.578     & 50  &17.34&  24.275 & 11 & 7  & 3 &15.00&0.6429$\pm$ 0.1430   \\
			\hline
		\end{tabular}
	\end{table*}
\end{center}

\begin{figure}[t]
	\centerline{\includegraphics[width=108mm,height=11.75pc]{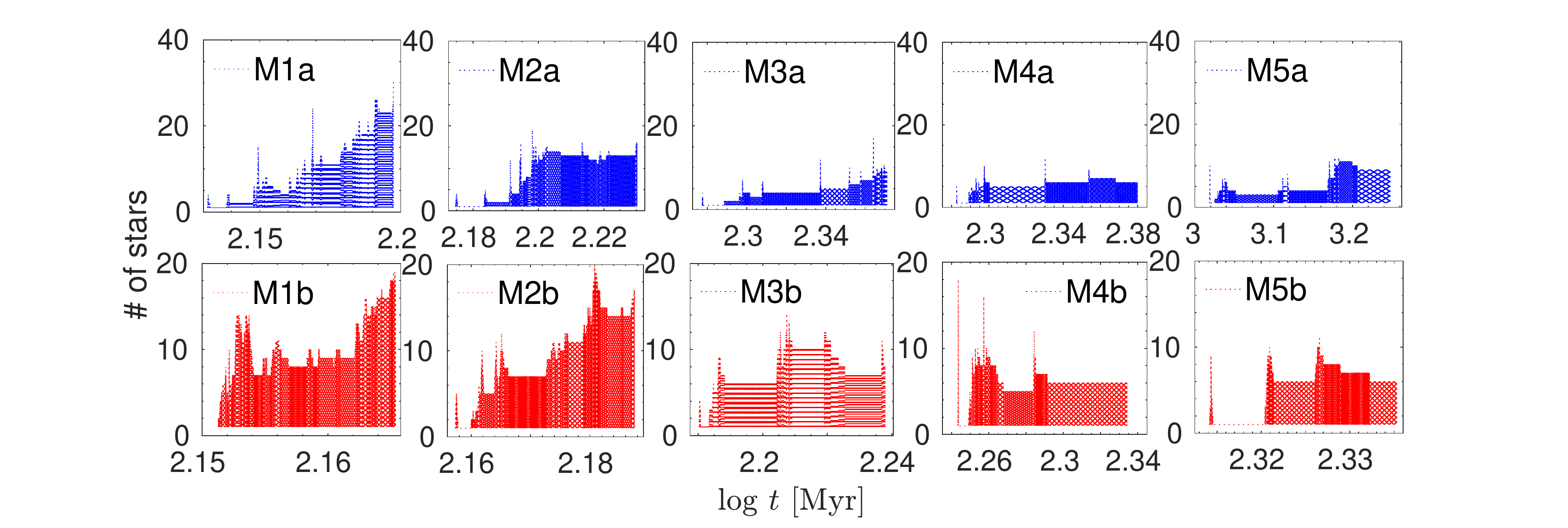}}
	\caption{Number of protostars as a function of time during the collapse of models M1a$-$M5a (top-panels) and models M1b$-$M5b (bottom-panels). All panels represent the point in time when the SFE $\xi$ reaches 15 \% in each model. The time is mentioned in Myr. Color in online edition.\label{Fig2}}
\end{figure}
Figure 3 is a representation of the mean mass $M_{\rm mean}$ of the protostars as a function of SFE in both sets of models. The panel on the top shows the model set M1a$-$M5a. To understand the results better we mark the specific value of the SFE with a vertical dotted black line where the transition from a fragmentation to accretion dominated regime takes place \citep{riaz2020fragmentation}. We find $\xi$ $\sim$ 5 \% as the transition point between these two regimes.  

In the top panel of Figure 3 we find that the mean values of the protostellar mass for the set of models M1a$-$M5a which is related to the thermal states in the star forming gas (see also Table 1) for $\xi$ $\sim$ 5 \%, providing evidence for higher average masses once accretion takes over in models with higher initial thermal state. The total number of fragments is also reduced when the initial temperature is higher. Similarly, the bottom panel in Figure 3 shows the second set of models M1b$-$M5b, which, in general, repeats the trend observed in models M1a$-$M5a. The point of transition from the fragmentation dominated regime to the accretion dominated regime in case of the second set of models appears at $\xi$ $\sim$ 5 \%. In the two sets of models we see that from $\xi$ = 5 \% to $\xi$ = 15 \% the warmer clouds yield higher mean masses $M_{\rm mean}$ than clouds with cold molecular gas. 

In Figure 4 we present the total mass accumulation history until the SFE reaches to $\xi$ = 15 \%. The gas clouds initiate their collapse from their respective floor temperatures which according to our assumption are primarily set by the CMB radiation. We observe that the mass of the gas cloud converting into protostars throughout the dynamical history remains on the rise for models which initiate from a lower floor temperature. However, the gas clouds collapsing from a higher floor temperature exhibit a rather smoothed mass accumulation history during the later part of their evolution. This is mainly due to the fact that cold gas environment, in general, triggers formation of greater number of protostars because of the prevailing conditions supporting smaller Jeans mass which subsequently via frequent merger events keep the total mass accumulation profile on a sharp rise until the SFE reaches to 15 \%. Contrary to this, warm gas cloud environment, in general, suppresses the formation of protostars and prevent greater number of gas clumps to appear as self-gravitating objects. Thus with a smaller number of protostars and hence with fewer merger events the protostars, in general, rely on the phenomenon of gas accretion to contribute towards the SFE. This feature makes the total mass accumulation profile to show a trend of smooth rise specially at the later part of these models evolution. 

\begin{figure}[t]
	\centerline{\includegraphics[width=75mm,height=20pc]{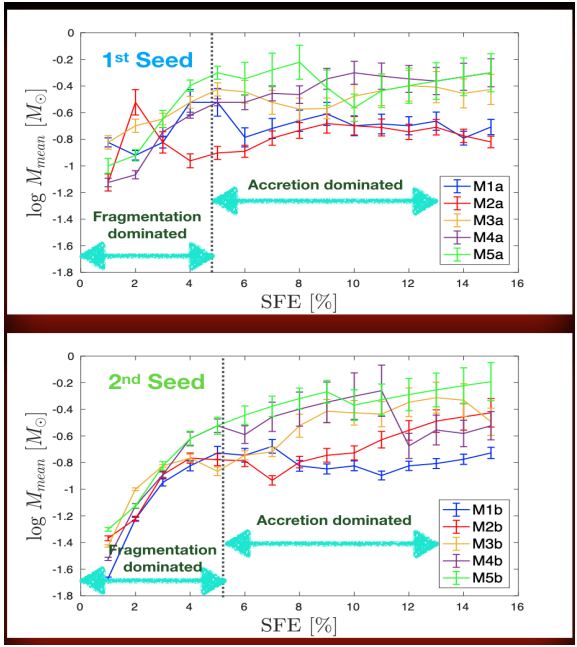}}
	\caption{Top - Mean mass of protostars $M_{\rm mean}$ as a function of SFE for set of models M1a$-$M5a. Bottom - Mean mass of protostars $M_{\rm mean}$ as a function of SFE for set of models M1b$-$M5b. The mean mass of protostars $M_{\rm mean}$ is given in units of solar mass. The dotted vertical line in each panel indicates the transition point between fragmentation dominated and accretion dominated regimes. These profiles represent the stage of evolution where the SFE in each model reaches $\xi$ = 15 \%. Color in online edition. \label{Fig3}}
\end{figure}

\begin{figure}[t]
	\centerline{\includegraphics[width=105mm,height=11pc]{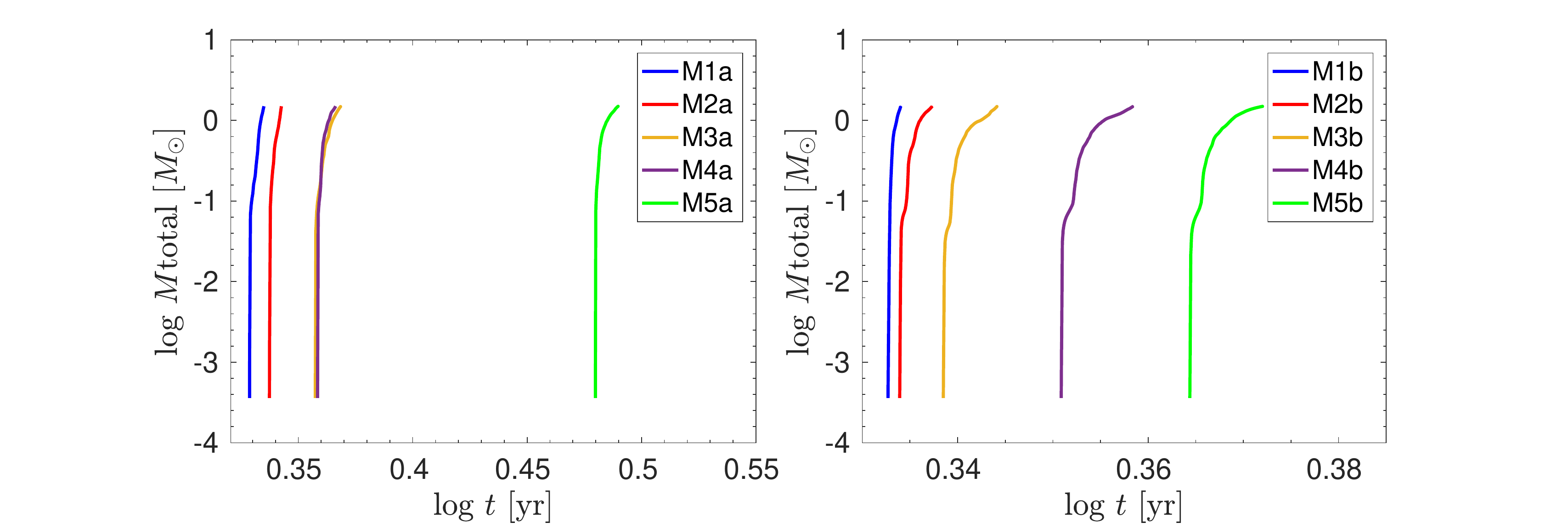}}
	\caption{The total mass accumulation by protostars as a function of time. Right - total mass of all protostars $M_{\rm total}$ set of models M1a$-$M5a. Left - Mean mass of protostars $M_{\rm total}$ for set of models M1b$-$M5b. The total mass of protostars $M_{\rm total}$ is given in units of solar mass. The time is given in Myrs. These profiles represent the stage of evolution where the SFE in each model reaches $\xi$ = 15 \%. Color in online edition.\label{Fig4}}
\end{figure}

\begin{figure}[t]
	\centerline{\includegraphics[width=110mm,height=10.25pc]{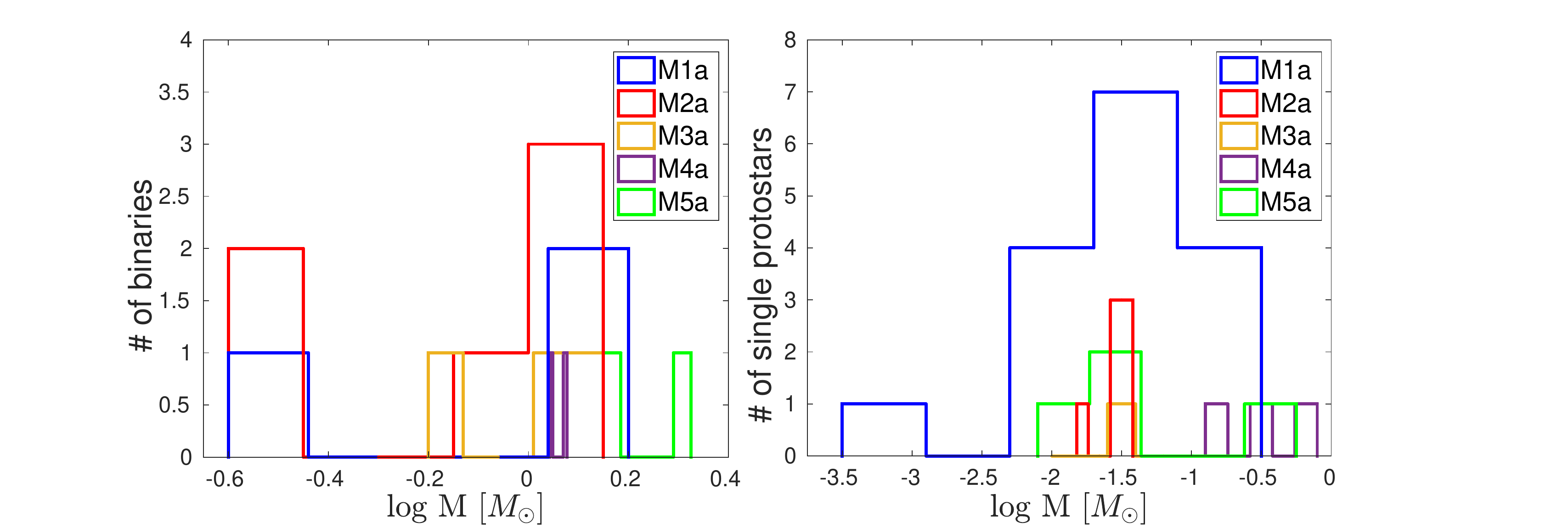}} \centerline{\includegraphics[width=110mm,height=10.25pc]{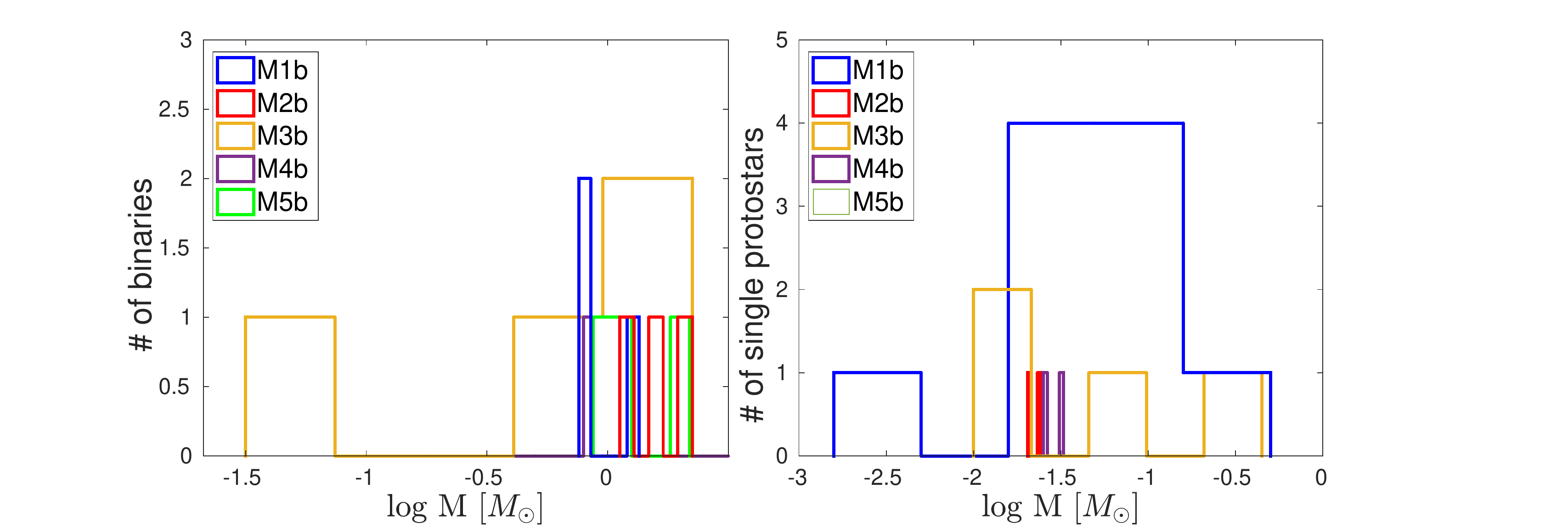}}
	\caption{Left (top \& bottom) - Number of binary protostars as a function of their mass for set of models M1a$-$M5a and M1b$-$M5b, respectively. Right (top \& bottom) - Number of isolated protostars as a function of their mass for set of models M1a$-$M5a and M1b$-$M5b, respectively. The mass of the binary and the isolated protostars is given in units of solar mass. These mass distributions represent the stage of evolution where the SFE in each model reaches $\xi$ = 15 \%. Color in online edition.\label{Fig4}}
\end{figure}

In Figures 5, we present for respective $set~1$ (top-panels) and $set~2$ (bottom panels), the total mass of the protostars falling inside the binary and the isolated stellar configurations. As found in the previous studies, the binary systems are the most favorable stellar configuration in most stellar populations \citep{goodwin2010binaries,duquennoy1991multiplicity,eggleton2008catalogue}. Most binaries are formed as binaries and not built by stellar encounters in the dynamical evolution of dense stellar systems \citep{goodman1993binary}. In our numerical experiment here we also form binaries in every simulation model (see Table 1). Due to the larger number of protostars present in the cold gas clouds than in the warm gas cloud, the number of binaries follows the similar trend. More massive binaries are formed in gas clouds that have initial higher thermal states. Whereas the colder gas clouds form binary systems with a wider range of possible protostellar masses. 

\section{Conclusions}\label{sec5}

Warmer gas clouds when going through gravitationally collapse do form massive protostars. In this process, the number of protostars remains limited due to the higher Jeans mass conditions prevailing inside the gas. Colder gas clouds typically yield lower-mass protostars. Although we observe more merger events in colder systems, they are not frequent enough to lead to similar numbers of high-mass objects as we see in the warmer clouds.

We have found that the stellar mean masses respond to the two main processes of gas fragmentation and mass accretion. Our mean mass analysis have shown that the gas fragmentation dominated regime has a weak dependence on the initial thermal state of the gas. In the accretion dominated regime, however, we find that the stellar means mass correlates with the gas temperature \citep{riaz2020fragmentation}.

Once higher SFEs are reached, our simulations show that one would expect a dependence of the accretion process on temperature. This could be caused by the local radiation backgrounds that heat up the gas and allow the protostars to accrete the surrounding material more efficiently.

Protobinary systems remain a feature of all simulations regardless of their initial conditions. As expected, warmer gas clouds produce more massive binary systems while the colder gas clouds, in general, form less massive binaries. The binary contribution towards the total SFE of $\xi$ $=$ 15 \% remains on much higher side when compared with the isolated protostars in our simulations. 

Our simulation results support a top-heavy IMF for the stellar population coming into existence at higher redshift environments where the CMB radiation regulates the thermal state of the star forming gas and affect the stellar mass.

\section*{Acknowledgments}
This research was supported by the supercomputing infrastructure of the NLHPC (ECM$-$02) and the Kultrun Astronomy Hybrid Cluster (projects ANID Programa de Astronomia Fondo Quimal QUIMAL 170001, ANID PIA ACT172033, and Fondecyt Iniciacion 11170268). RR and the second author DRGS thank for funding through Fondecyt Postdoctorado (project code 3190344) and the Concurso Proyectos Internacionales de Investigaci\'on, Convocatoria 2015" (project code PII20150171). DRGS further thanks for funding via Fondecyt regular (project code 1161247) and via the Chilean BASAL Centro de Excelencia en Astrof\'isica yTecnolog\'ias Afines (CATA) grant PFB-06/2007. SV thanks to Prof. Dr. R. Keppens and Prof. Dr. S. Poedts for the KUL supercomputing cluster Thinking. RSK acknowledges financial support from the German Research Foundation (DFG) via the collaborative research centre (SFB 881, Project-ID 138713538) ``The Milky Way System'' (subprojects B1, B2, and B8) and from the Heidelberg cluster of excellence EXC 2181 (Project-ID 390900948) ``STRUCTURES: A unifying approach to emergent phenomena in the physical world, mathematics, and complex data'' funded by the German Excellence Strategy.
 
\bibliography{Riaz}%

\end{document}